\documentstyle[11pt]{article}

\begin{document} 
\centerline{\bf Numerical simulation of a Controlled-Controlled-Not (CCN) quantum gate }
\centerline{\bf in a chain of three interacting nuclear spins system}
\vskip2pc
\centerline{G.V. L\'opez and L. Lara}
\centerline{Departamento de F\'{\i}sica, Universidad de Guadalajara}
\centerline{Apartado Postal  4-137, 44410 Guadalajara, Jalisco, M\'exico}
\vskip2pc
\centerline{PACS: 03.67.Lx, 03.65.Ta}
\vskip2cm
\centerline{ABSTRACT}
\vskip1pc\noindent
We present the study of a quantum Controlled-Controlled-Not gate, implemented in a chain of three nuclear
spins weakly Ising interacting between all of them, that is, taking into account first and second
neighbor spin interactions. This implementation is done using a single resonant
$\pi$-pulse on the initial state of the system (digital and superposition). The fidelity parameter is
used to determine the behavior of the CCN quantum gate as a function of the ratio of the second neighbor
interaction coupling constant to the first neighbor interaction coupling constant  ($J'/J$). We found
that for $J'/J\ge 0.02$ we can have a well defined CCN quantum gate.
\vfil\eject\noindent 
{\bf 1. Introduction}
\vskip0.5pc\noindent
There is not doubt that the discovery of the polynomial time solution of the prime factorization problem
(Shor's algorithm [1]) and the fastest data base searching (Grover's algorithm [2]) by a quantum
computer, and the importance of the implementations of these algorithms in cryptography analysis [3]
have made the study of the physical realization of a quantum computer a priority for many researcher
[4,5]. A quantum computer works with qubits instead of bit, where a qubit is the superposition of two
quantum levels of the system which are defined as $|0\rangle$ and $|1\rangle$, and a tensorial product of
these states makes up a register. Although, one could say that there are already quantum computers with
registers of few qubits [6], these are still far too low to make serious studies (which may require at
least 100-qubits registers, that is, at least a $2^{100}$ dimensional Hilbert space). In
addition, Much more technology development (or very neat idea) is needed to have a real powerful quantum
computer. One model of solid state quantum computer which looks promising due to advances in  single spin
measurements [7] is that of a chain of nuclear spins quantum computer
[4,8], where the Ising interaction between first neighbor is considered. This interaction allows us to
implement ideally this type of computer up to 1000 qubits or more [9]. First neighbor interaction
between spins also allows to implement a Controlled-Not (CN) quantum gate using a single $\pi$-pulse
[10]. However, the implementation of a Controlled-Controlled-Not (CCN) quantum gate, which is needed
in Shor's factorization, Grover's searching and Full-Adder algorithms [11], is not that simple. In
addition, CCN quantum gate is also important since it has universality characteristic [12]. The CN and
CCN gates are defined classically by the following two tables [13] 
\vskip1pc
\centerline{\vbox{\tabskip=0pt \offinterlineskip
\def\tablerule{\noalign{\hrule}}
\halign to125pt{\strut#& 
               \vrule#\tabskip=1em plus5em&
  \hfil#\hfil&      \vrule#& 
  \hfil#\hfil&      \vrule#&
  \hfil#\hfil&      \vrule#&
  \hfil#\hfil&      \vrule#&
  \hfil#\hfil&      \vrule#
\tabskip=0pt\cr\tablerule
&&\multispan7\hfil CN\hfil&\cr\tablerule
&&\omit\hidewidth a  \hidewidth
&&\omit\hidewidth b  \hidewidth
&&\omit\hidewidth a' \hidewidth
&&\omit\hidewidth b' \hidewidth&\cr\tablerule
&& 0 && 0 && 0 && 0 &\cr\tablerule
&& 0 && 1 && 0 && 1 &\cr\tablerule
&& 1 && 0 && 1 && 1 &\cr\tablerule
&& 1 && 1 && 1 && 0 &\cr\tablerule \noalign{\smallskip}
&\multispan7\hfil\cr}} } 
\centerline{ Table 1}
\vskip2pc
\centerline{
\vbox{\tabskip=0pt \offinterlineskip
\def\tablerule{\noalign{\hrule}}
\halign to189pt{\strut#& 
               \vrule#\tabskip=1em plus5em&
  \hfil#\hfil&      \vrule#& 
  \hfil#\hfil&      \vrule#&
  \hfil#\hfil&      \vrule#&
  \hfil#\hfil&      \vrule#&
  \hfil#\hfil&      \vrule#&
  \hfil#\hfil&      \vrule#& 
  \hfil#\hfil&      \vrule#& 
  \hfil#\hfil&      \vrule#&
  \hfil#\hfil&      \vrule#&
  \hfil#\hfil&      \vrule#&
  \hfil#\hfil&      \vrule#
\tabskip=0pt\cr\tablerule
&&\multispan{11}\hfil CCN\hfil&\cr\tablerule
&&\omit\hidewidth a  \hidewidth
&&\omit\hidewidth b  \hidewidth
&&\omit\hidewidth c  \hidewidth
&&\omit\hidewidth a' \hidewidth
&&\omit\hidewidth b' \hidewidth
&&\omit\hidewidth c' \hidewidth&\cr\tablerule
&& 0 && 0 && 0 && 0 && 0 && 0 &\cr\tablerule
&& 0 && 0 && 1 && 0 && 0 && 1 &\cr\tablerule
&& 0 && 1 && 0 && 0 && 1 && 0 &\cr\tablerule
&& 0 && 1 && 1 && 0 && 1 && 1 &\cr\tablerule
&& 1 && 0 && 0 && 1 && 0 && 0 &\cr\tablerule
&& 1 && 0 && 1 && 1 && 0 && 1 &\cr\tablerule
&& 1 && 1 && 0 && 1 && 1 && 1 &\cr\tablerule 
&& 1 && 1 && 1 && 1 && 1 && 0 &\cr\tablerule
\noalign{\smallskip}
&\multispan9\hfil\cr}} }
\centerline{Table 2}
\vskip1pc\noindent
where a (called control bit) and b (called target bit) for CN (a,b and c for CCN) represent the input
information, and a' and b' (a', b', and c' for CNN) represent the output information. In a CN gate, "b'"
(target) changes if an only if "a"(control) has the value 1. In a CCN gate, "c'" changes if an only if
"a" and "b" has the value 1. The ideals CN and CCN quantum gates, operating on an arbitrary
L-register
$|i_{L-1},\dots,i_a,\dots,i_b,\dots,i_c,\dots,i_0\rangle$ (where $i_j=0,1$), would be defined as
$$CN_{ab}|i_{L-1},\dots,i_a,\dots,i_b,\dots,i_c,\dots,i_0\rangle=
|i_{L-1},\dots,i_a,\dots,i_b\oplus i_a,\dots,i_c,\dots,i_0\rangle$$
and
$$CCN_{abc}|i_{L-1},\dots,i_a,\dots,i_b,\dots,i_c,\dots,i_0\rangle=
|i_{L-1},\dots,i_a,\dots,i_b,\dots,i_c\oplus (i_a\cdot i_b),\dots,i_0\rangle\ ,$$
where the operation $\oplus$ means summation module 2. In this paper we
show that considering second neighbor interaction between spins in the one-dimensional chain of spins
quantum computer, and  we show that it is
possible to implement a CCN quantum gate with just a single $\pi$-pulse. Of course, a CN quantum gate
continue being represented by a single $\pi$-pulse.
\vskip2pc
\leftline{\bf 2. Equation of Motion}
\vskip1pc\noindent
Consider a one-dimensional chain of three equally spaced nuclear-spins system (spin one half) making an
angle $\cos\theta=1/\sqrt{3}$ with respect the z-component of the magnetic field (chosen in this way to
kill the dipole-dipole interaction between spins) and having an rf-magnetic field in the transversal
plane. The magnetic field is given by
$${\bf B}=(b\cos\omega t, -b \sin\omega t, B(z))\ ,\eqno(1)$$
where $b$  is the amplitude of the rf-field, $B(z)$ is the amplitude of the z-component of the magnetic
field, $\omega$ is the angular frequency of the rf-field (its phase has been chosen as zero for
simplicity). So, the Hamiltonian of the system is given by
$$H=-\sum_{k=0}^2{\bf \mu_k}\cdot {\bf B_k}-2J\hbar\sum_{k=0}^1I_k^zI_{k+1}^z
-2J'\hbar\sum_{k=0}^0I_k^zI_{k+2}^z\ ,\eqno(2)$$
where ${\bf\mu_k}$ represents the magnetic moment of the kth-nucleus which is given in terms of the
nuclear spin as ${\bf\mu_k}=\hbar\gamma(I_k^x,I_k^y, I_k^z)$, being $\gamma$ the proton gyromagnetic
ratio. ${\bf B_k}$ represents the magnetic field at the location of the $kth$-spin. The second term at the
right side of (2)  represents the  first neighbor spin interaction, and the third term represents the
second neighbor spin interaction. $J$ and $J'$ are the coupling constants for these interactions. This
Hamiltonian can be written in the  following way
$$H=H_0+W\ ,\eqno(3a)$$
where $H_0$ and $W$ are given by
$$H_0=-\hbar\left\{\sum_{k=0}^2\omega_kI_k^z+2J(I_0^zI_1^z+I_1^zI_2^z)+2J'I_0^zI_2^z\right\}\eqno(3b)$$
and
$$W=-{\hbar\Omega\over 2}\sum_{k=0}^2\biggl[e^{i\omega t}I_k^++e^{-i\omega t}I_k^-\biggr]\ .\eqno(3c)$$
Here, $\omega_k=\gamma B(z_k)$ is the Larmore frequency of the kth-spin, $\Omega=\gamma b$ is the Rabi's
frequency, and $I_k^{\pm}=I_k^x\pm iI_k^y$ represents the ascend operator (+) or the descend operator (-).
The Hamiltonian $H_0$ is diagonal on the basis $\{|i_2i_1i_0\rangle\}$ with $i_j=0,1$ (zero for the
ground state and one for the exited state),
$$H_0|i_2i_1i_0\rangle=E_{i_2i_1i_0}|i_2i_1i_0\rangle\ .\eqno(4a)$$
The eigenvalues $E_{i_2i_1i_0}$ are given by
$$E_{i_2i_1i_0}=-{\hbar\over 2}\biggl\{(-1)^{i_2}\omega_2+(-)^{i_1}\omega_1+(-1)^{i_0}\omega_0+
J[(-1)^{i_0+i_1}+(-1)^{i_1+i_2}]+(-1)^{i_0+i_2}J'\biggr\}\ .\eqno(4b)$$
The term (3c) of the Hamiltonian (3a) allows to have a single spin transitions on the above eigenstates
by choosing the proper resonant frequency, as shown in Figure 1. For example, if we are  interested in
having the transition $|110\rangle\longleftrightarrow |111\rangle$ since this one represents the CCN
quantum gate operation, according with the table 2, the resonant frequency would be
$$\omega=\omega_0-J-J'\ .\eqno(5)$$
Of course, we could also have considered a CCN quantum gate as defined by the transition $|011\rangle
\longleftrightarrow |111\rangle$, by defining the order of the elements differently. 
\vskip1pc\noindent
To solve the Schr\"odinger equation
$$i\hbar{\partial\Psi\over\partial t}=H\Psi\ ,\eqno(6)$$ 
let us propose a solution of the form
$$\Psi(t)=\sum_{k=0}^7C_k(t)|k\rangle\ ,\eqno(6)$$
where we have used decimal notation for the eigenstates in (4a), $H_0|k\rangle=E_k|k\rangle$.
Substituting (6) in (5), multiplying for the bra $\langle m|$, and using the orthogonality relation 
$\langle m|k\rangle=\delta_{mk}$, we get the following equation for the coefficients
$$i\hbar\dot C_m=E_mC_m+\sum_{k=0}^7C_k\langle m|W|k\rangle\ \ m=0,\dots,7.\eqno(7)$$
Now, using the following transformation
$$C_m=D_me^{-iE_m t/\hbar}\ ,\eqno(8)$$
the fast oscillation term $E_mC_m$ of Eq. (7) is removed (this is equivalent to going to the rotating
frame of reference), and the following equation is gotten for the coefficients $D_m$
$$i\dot D_m={1\over\hbar}\sum_{k=0}^7W_{mk}D_ke^{i\omega_{mk}t}\ ,\eqno(9a)$$
where $W_{mk}$  denotes the matrix elements $\langle m|W|k\rangle$, and $\omega_{mk}$ are defined as
$$\omega_{mk}={E_m-E_k\over\hbar}\ .\eqno(9b)$$
Eq. (9a) represents a set  of sixteen real coupling ordinary differential equations which can be solved
numerically, and where $W_{mk}$ are the elements of the matrix
$$(W)=-{\hbar\Omega\over 2}\pmatrix{
0 & z^* & z^* & 0   & z^* & 0   & 0   & 0 \cr
z &  0  &  0  & z^* & 0   & z^* & 0   & 0 \cr
z &  0  &  0  & z^* & 0   & 0   & z^* & 0 \cr
0 &  z  &  z  &  0  & 0   & 0   & 0   & z^*\cr
z &  0  &  0  &  0  &  0  & z^* & z^* & 0  \cr
0 &  z  &  0  &  0  & z   &  0  & 0   & z^*\cr
0 &  0  &  z  &  0  &  z  &  0  & 0   & z^*\cr
0 &  0  &  0  &  z  &  0  &  z  & z   & 0  \cr}\ ,\eqno(9c)$$
where $z$ is defined as $z=e^{i\omega t}$, and $z^*$ is its complex conjugated.
\vskip2pc
\leftline{\bf 3. Numerical Simulations}
\vskip1pc\noindent
To solve numerically (9a), we shall use similar values for the parameters as reference (9). So, in units
of $2\pi\times MHz$, we set the following values
$$\omega_0=100\ ,\ \omega_1=200\ ,\ \omega_2=400\ ,\ J=5\ ,\ \Omega=0.1\eqno(10)$$
The coupling constant $J'$ is chosen with at least one order of magnitude less that $J$ since in the
chain of spins one expects that second neighbor contribution to be at least one order of magnitude weaker
than first neighbor contribution. We consider digital initial state and superposition initial
states,
$$\Psi(0)=|110\rangle\eqno(10a)$$
and
\begin{eqnarray*}
\Psi(0)&=&{2\over 3\sqrt{8}}|000\rangle+{\sqrt{14}\over 3\sqrt{8}}|001\rangle+{1\over
3\sqrt{8}}|010\rangle+{\sqrt{17}\over 3\sqrt{8}}|011\rangle\\
& &+{3\over 4\sqrt{8}}|100\rangle+
{\sqrt{23}\over 4\sqrt{8}}|101\rangle+{1\over 2\sqrt{8}}|110\rangle+{\sqrt{7}\over 2\sqrt{8}}|111\rangle
\ ,
\end{eqnarray*}
$$\eqno(10b)$$
to cover all the important aspects of the CNN quantum gate. In all our simulations the total probability,
$\sum|C_k(t)|^2$, is conserved equal to one within a precision of $10^{-6}$ (note that $|C_k|=|D_k|$).
Figure 2 shows the behavior of $Re~ D_6$, 
$Im~D_6$, $Re~D_7$ and $Im~D_7$  during a $\pi$-pulse ($t=\tau=\pi/\Omega$) for the digital initial
state and with $J'=0.1$. One can see clearly the transition $|110\rangle\longrightarrow i|111\rangle$, 
defining the CCN quantum gate up to a global phase ($\pi/2$). Figure 3a shows the behavior of the
probabilities $|C_k|$ for $k=0,\dots,7$ during a $\pi$-pulse for the superposition initial state and with
$J'=0.1$ and for the superposition initial condition. Figure 3b shows the behavior of the expected value
of the z-component of the spin,
$\langle I_k^z\rangle$, k=0,1,2, during a $\pi$-pulse with the initial superposition state and $J'=0.1$.
These expected values are given by
$$\langle I_0^z\rangle={1\over 2}\sum_{k=0}^7(-1)^k|C_k(t)|^2\ ,\eqno(11a)$$
$$\langle I_1^z\rangle={1\over
2}\biggl\{|C_0|^2+|C_1|^2-|C_2|^2-|C_3|^2+|C_4|^2+|C_5|^2-|C_6|^2-|C_7|^2\biggr\}\ ,\eqno(11b)$$
and
$$\langle I_2^z\rangle={1\over 2}\sum_{k=0}^3|C_k|^2-\sum_{k=4}^7|C_k|^2\ .\eqno(11c)$$
As one could expect, only the spin related with the first qubit (having subindex zero) changes its value
at the end of the
$\pi$-pulse. The behavior of the expected values of the $x$ and $y$ components of the spin,
$\langle I_k^x\rangle$ and $\langle I_k^y\rangle$ for $k=0,1,2$, is shown on Figure 4 for the
superposition initial state and for $j'=0.1$. These expected values are given by
$$\langle I_0^x\rangle=Re\biggl(C_1^*C_0+C_3^*C_2+C_5^*C_4+C^*_7C_6\biggr)\ ,\hskip1cm\langle
I_0^y\rangle=Im\biggl(\dots\biggr)\ ,\eqno(12a)$$
$$\langle I_1^x\rangle=Re\biggl(C_2^*C_0+C_3^*C_1+C_6^*C_4+C^*_7C_5\biggr)\ ,\hskip1cm\langle
I_1^y\rangle=Im\biggl(\dots\biggr)\ ,\eqno(12b)$$ 
and
$$\langle I_2^x\rangle=Re\biggl(C_4^*C_0+C_5^*C_1+C_6^*C_2+C^*_7C_3\biggr)\ ,\hskip1cm\langle
I_2^y\rangle=Im\biggl(\dots\biggr)\ .\eqno(12c)$$
The fast oscillations appearing on this behavior are due to the time depending phases in Eq. (8) which
have these terms and which are not cancelled out.
\vskip0.5pc\noindent
For the case $J'=0$, the resonant frequency for the transition $|110\rangle\longleftrightarrow
|111\rangle$ and the transition $|010\rangle\longleftrightarrow |011\rangle$ coincides and both
transitions appears in the dynamics of the system (simultaneous CN and CCN operations). Thus, the
question which arises is: what is the minimum value of $J'$ which allows to have a well define CCN
quantum gate ? To answer this question, Fig. 5 (a) and (b) shows the behavior of the CN quantum gate with
the resonant frequency of the CCN quantum gate ($\omega=\omega_0-J-J'$) during a $\pi$-pulse for several
$J'$ values. In addition, (c) and (d) show the fidelity [14],
$F=\langle\Psi_{expected}|\Psi(\tau)\rangle$, associated to CCN quantum gate as a function of $J'/J$. As
one can see from these figures, with $J'$ values of even two orders of magnitude ($J'/J=0.02$) we can
already have a very well defined CCN quantum gate. Therefore, it is very likely that a single pulse CCN
quantum gate can be implemented in a chain of nuclear spin quantum computer.
\vskip3pc\noindent
\leftline{\bf 4. Conclusions and comments}
\vskip0.5pc\noindent
 We presented the study of a Controlled-Controlled-Not (CCN) quantum gate implemented in a chain of
nuclear spins weakly Ising interacting and with the use of a single $\pi$-pulse. With study this gate
with digital and superposition initial states and found an expected global phase for its definition,
\begin{eqnarray*}
\widehat{ CCN}&=&|000\rangle\langle 000|+|001\rangle\langle 001|+|010\rangle\langle 010|+
|011\rangle\langle 011|+|100\rangle\langle 100| \\
&+&|101\rangle\langle 101| + i|110\rangle\langle 111|+i|111\rangle\langle 110|\ .
\end{eqnarray*}
$$\eqno(13)$$
Using the fidelity parameter, we have seen that it is very likely to have the implementation of a CCN
quantum gate in the chain of nuclear spin quantum computer since the coupling constant to second neighbor
interaction can be even two order of magnitude lower than the first neighbor interaction, and still having
a well defined CCN quantum gate. Since the eigenvalues in (4a) depends on $J'$, Eq. (4b), the detuning
factor, defined as $\Delta=(E_p-E_m)/\hbar-\omega$ for the transition
$|p\rangle\longleftrightarrow|m\rangle$ (decimal notation), will depends also on
$J'$. Therefore, the normal
$2\pi$k-method used in reference (8) will have to change correspondently to be able to suppress
non-resonant transitions in multiple-pulses algorithms.
\vskip5pc\noindent
\leftline{\bf Acknowledgements}
\vskip0.5pc\noindent
 This work was supported by SEP under the contract PROMEP/103.5/04/1911 and the University of Guadalajara.
\vfil\eject
\leftline{\bf Figure Captions}
\vskip1pc\noindent
Fig. 1 Energy levels and resonant frequencies.
\vskip0.5pc\noindent
Fig. 2 CCN quantum gate with digital initial condition and with $J'=0.1$. (1):$Re~D_6$, (2):$Im~D_6$,
(3):$Re~D_7$, (4):$Im~D_7$.
\vskip0.5pc\noindent
Fig. 3 For the superposition initial state and for $J'=0.1$, (a) Probabilities $|C_k(t)|^2$ for
$k=0,\dots,7$. (b) Expected values (0): $\langle I_0^z\rangle$,  (1): $\langle I_1^z\rangle$, and
(2): $\langle I_2^z\rangle$.
\vskip0.5pc\noindent
Fig. 4 For the superposition initial state and for $J'=0.1$, expected values of the transversal
components of the spin.
\vskip0.5pc\noindent
Fig. 5 For the superposition initial condition. (a) Probability $|C_2|^2$, and (b) Probability
$|C_3|^2$ for (1): $J'=0.0$, (2): $J'=0.02$, (3): $J'=0.04$, (4): $J'=0.06$, (5): $J'=0.08$, and
(6): $J'=0.1$. (c) Real and Imaginary parts of the Fidelity. (c) Modulus of Fidelity. 
\vfil\eject
\leftline{\bf References}
\obeylines{
1. P.W. Shor, {\it Proc. of the 35th Annual Symposium on the Foundation
\quad of the Computer Science}, IEEE, Computer Society Press, N.Y. 1994, 124.
\quad P.W. Shor, Phys. Rev. A {\bf 52} (1995) R2493.
2. L.K. Grover, Phys. Rev. Lett., {\bf 79} (1997) 627.
\quad L.K. Grover, Science, {\bf 280} (1998) 228.
3. S. Singh, {\it The Code Book}, Anchor Books, 
\quad A Division of Random House, Inc., N.Y., 2000.
4. S. Lloyd, Science, {\bf 261} (1993) 1569.
5. C.H. Bennett, Physics Today, {\bf 48} (1995) 24.
\quad D.P. DiVincenzo, Science, {\bf 270} (1995) 255.
\quad N.A. Gershenfeld and J.L.Chuang, Science, {\bf 1997} 350.
\quad A. Steane, Rep. Prog. Phys., {\bf 61} (1998) 117.
\quad B.E. Kane, Nature, {\bf 393} (1998) 133.
\quad E. Knill, R. Laflamme, W.H. Zurek, Science, {\bf 279} (1998) 342.
6. L.M.K. Vandersypen, M. Steffen, G. Breyta, C.S. Yannomi,
\quad M.H. Sherwood and I.L. Chuang, Nature, {\bf 414} (2001) 883.
\quad F. Schmidt-Kaler, H.H\"affner, M. Riebe, G.P.T. Landcaster,
\quad T. Deuschle, C. Becher, W. H\"ansel, J. Eschner, C.F. Russ,
\quad and R. Blatt, Nature, {\bf 422} (2003) 408.
7. D. Rugar, R. Budakian, H.J. Manin and B.W. Chui, 
\quad Nature, {\bf 430}  (2004) 329.  
\quad P.C. Hammel, Nature, {\bf 430} (2004) 300.
\quad C. Hory and A.O. Hero, quant-ph/0402181.
8. G.P. Berman, G.D. Doolen, D.D. Holm and V.I. Tsifrinovich,
\quad Phys. Lett. A {\bf 193} (1994) 444.
9. G.P. Berman, G.D. Doolen, D.I. Kamenev, G.V. L\'opez, and
\quad  V.I. Tsifrinovich, Phys. Rev. A {\bf 6106} (2000) 2305.
10. G.P. Berman, D.K. Campbell, G.D. Doolen, G.V. L\'opez and 
\quad V.I. Tsifrinovich, Physica B {\bf 240} (1997) 61.
11. D. Beckman, A.N. Chari, S. Devabhaktuni, and I. Preskill,
\quad Phys. Rev. A, {\bf 54} (1996) 1034.
12. S. Lloyd, Phys. Rev. Lett., {\bf 75} (1995) 346.
13. R.P. Feynman, {\it Feynman Lectures on Computation},
\quad Addison-Wesley Publishing Co.,Inc., MA, 1996.
\quad R.P. Feynman, Found. of Phys., {\bf 16}, no. 6 (1986) 507.
14. A. Peres, Phys. Rev. A {\bf 30} (1984) 1610.
}

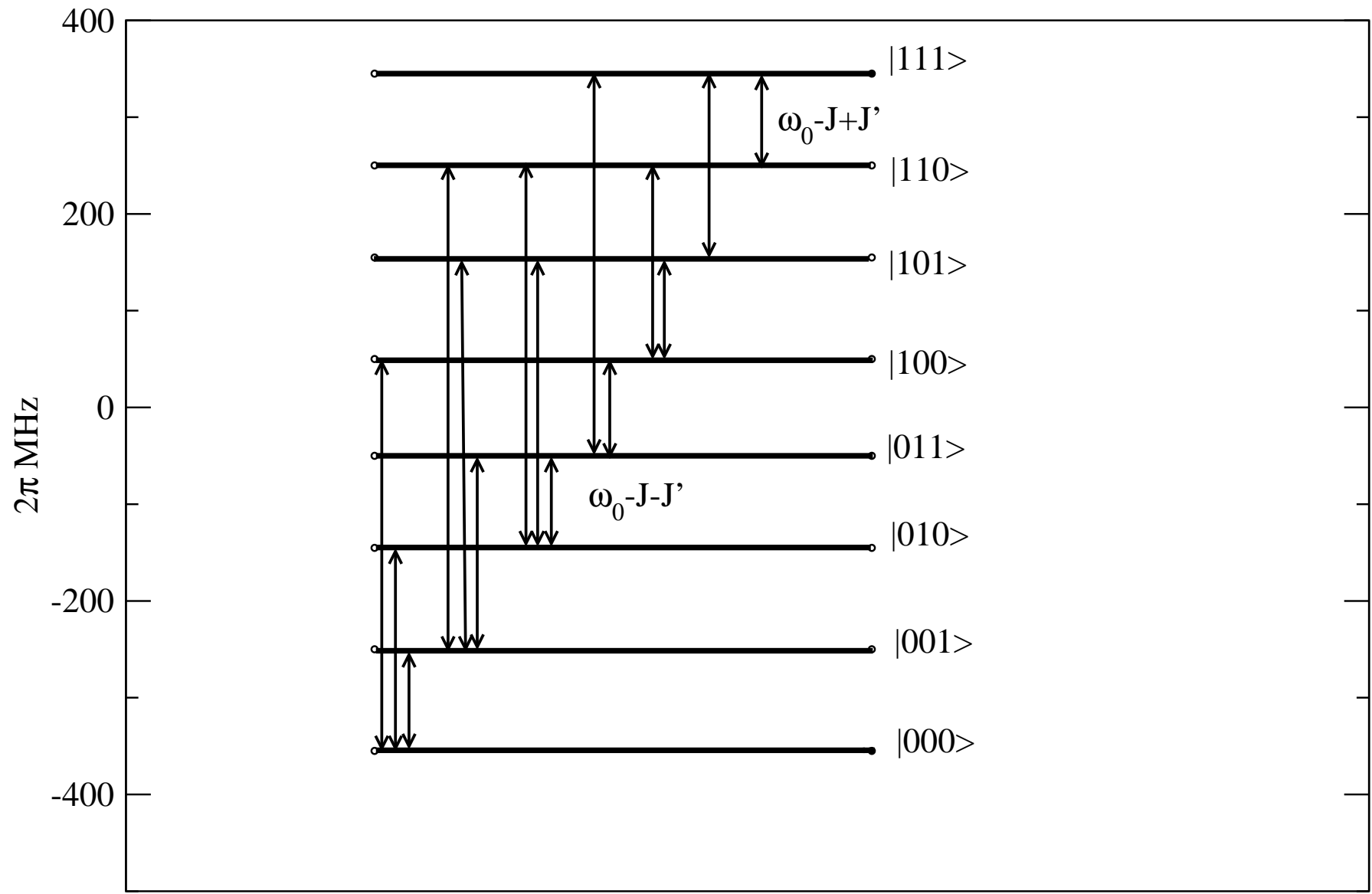
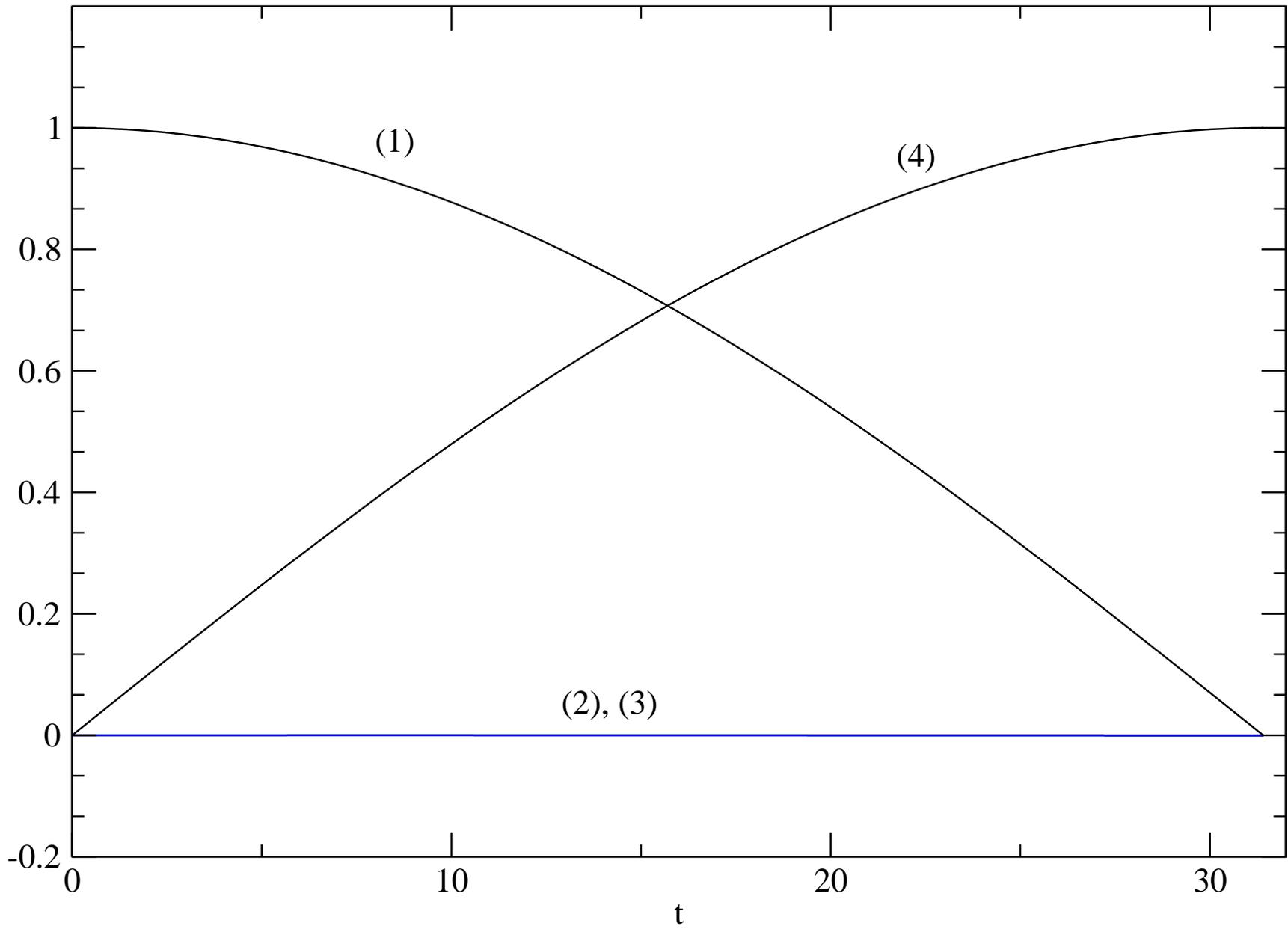
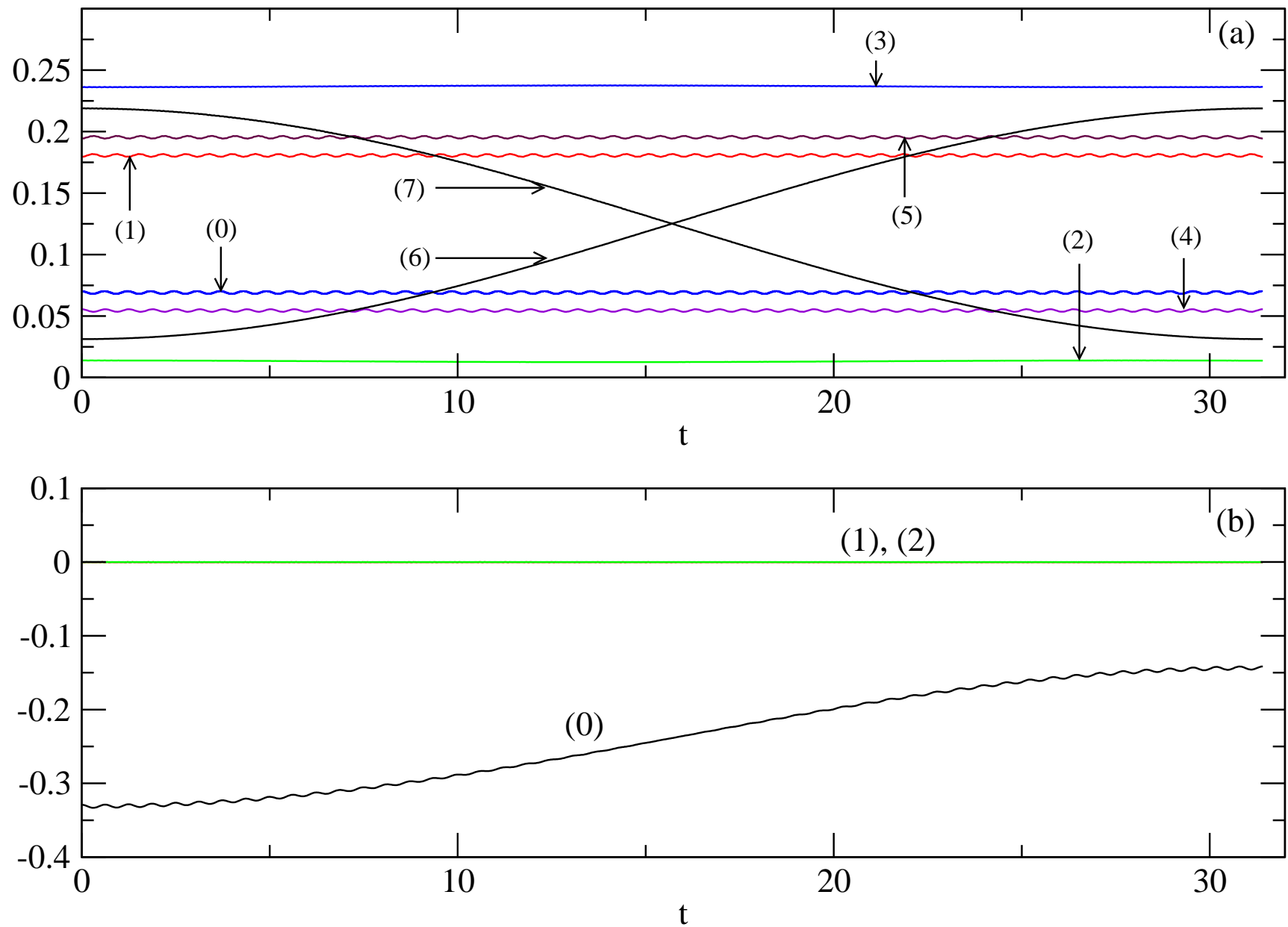
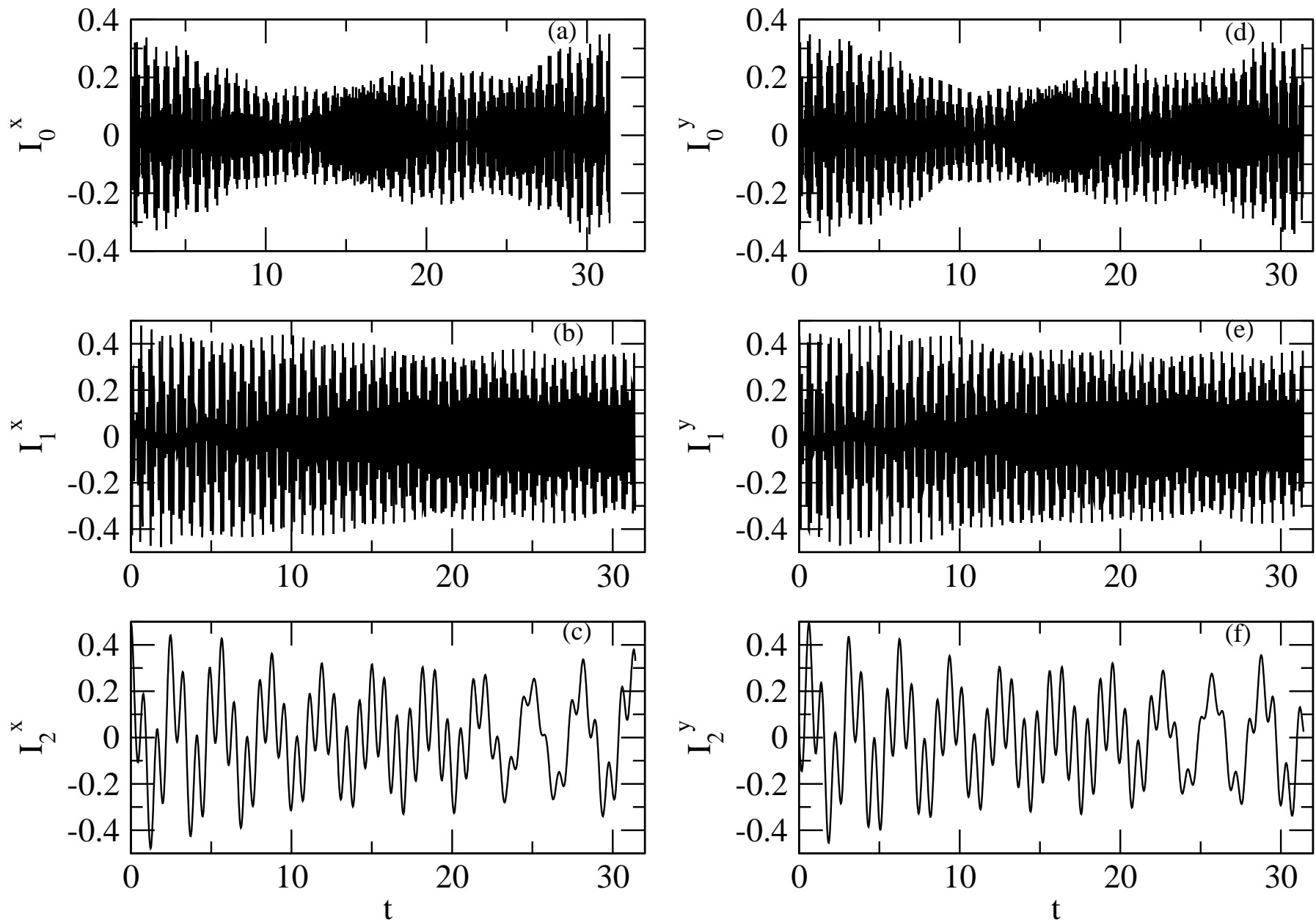
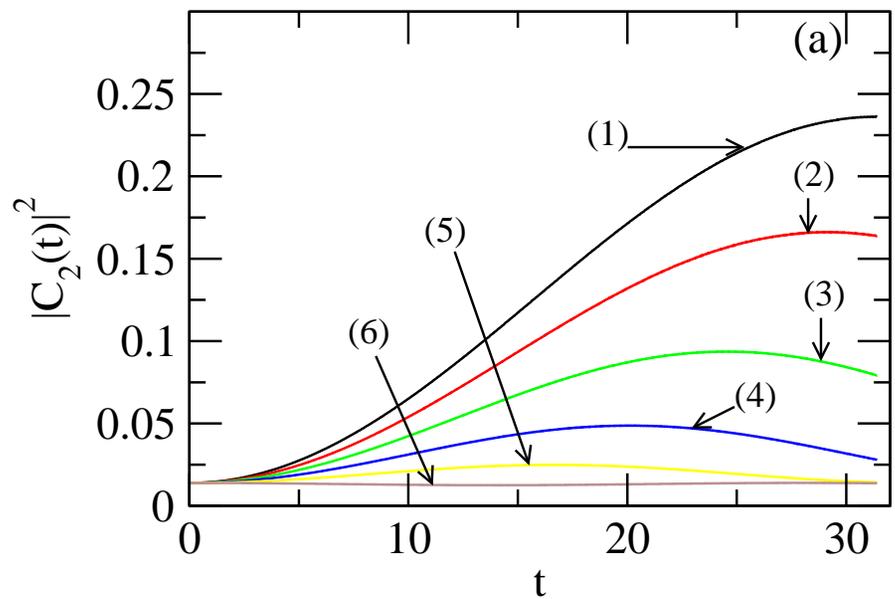
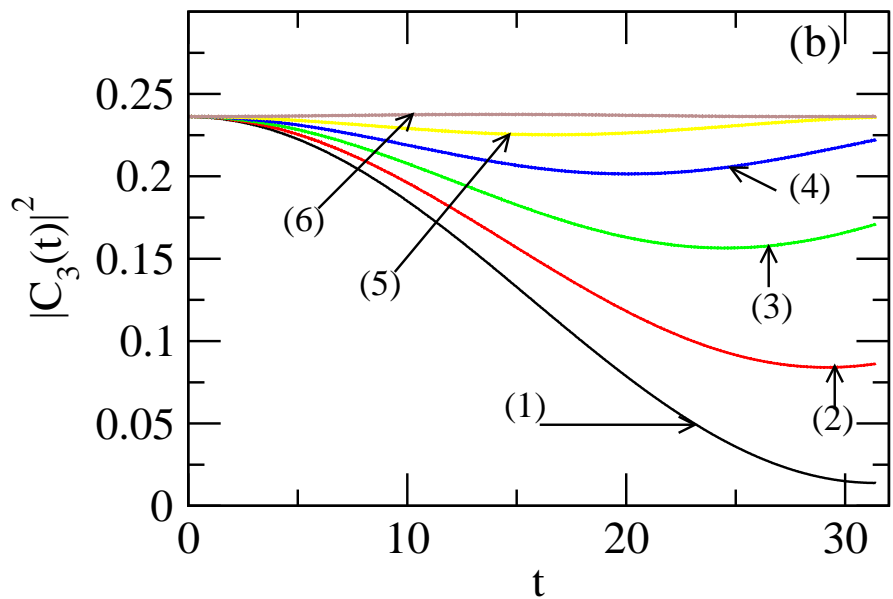
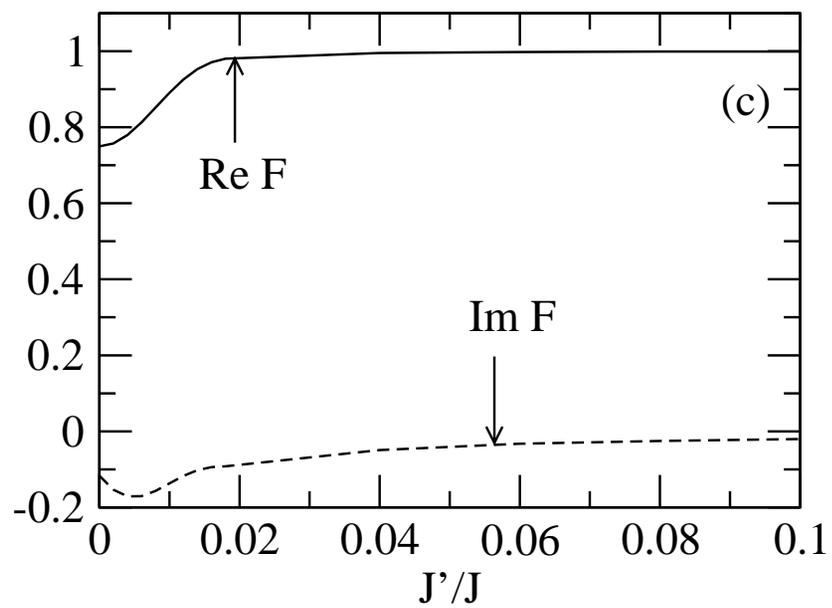
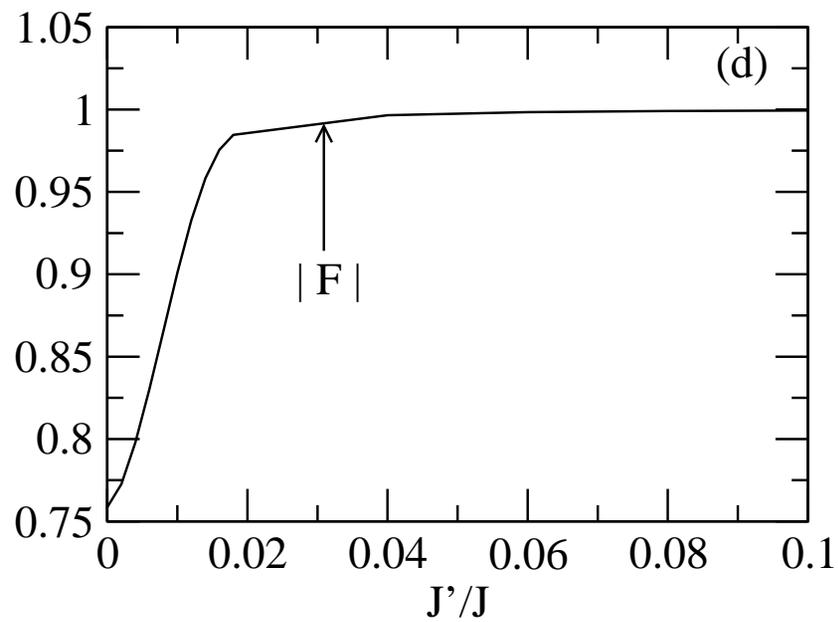
\end{document}